\documentstyle[seceq,preprint]{jpsj}

\def\runtitle{Mean Field Theory for the Bilayer Heisenberg Model}
\def\runauthor{Yasuhiro {\sc Matsushita},
Martin P. {\sc Gelfand}$^{1}$ and Chikara {\sc Ishii}}

\title
{\bf   Bond-Operator Mean Field Theory for the
Bilayer Heisenberg Model
} 
\author
{ Yasuhiro {\sc Matsushita}\footnote{E-mail: 
matusita@grad.ap.kagu.sut.ac.jp}, Martin P. {\sc Gelfand}$^{1}$ 
and Chikara {\sc Ishii}}

\inst
{
Department of Physics, Faculty of Science,
Science University of Tokyo, \\
Shinjuku-ku, Tokyo 162 \\
$^1$Department of Physics, 
Colorado State University, \\
Fort Collins, Colorado 80523, USA
}

\recdate
{
}

\abst
{ 
Bond-operator mean field equations for the square-lattice,
$S=1/2$ bilayer Heisenberg model are developed and solved
numerically.  
In the vicinity of both the zero-field
critical point and the field-induced transitions,
comparisons are made with $T=0$ and finite-temperature
strong coupling expansions.
The mean-field theory suggests that the quantum critical
region for the field-induced transitions is restricted to
significantly lower temperatures than one might have 
concluded based on strong-coupling expansions or other
numerical studies.
}
\kword
{bilayer Heisenberg model, 
field-induced transition, 
quantum critical phenomena, bond-operator mean-field 
theory, strong-coupling expansion} 

\begin{document}
\sloppy
\maketitle

\section{Introduction}
The appearances of unconventional superconductivity and
non-Fermi-liquid behavior near the quantum critical point 
of antiferromagnetic instability in strongly correlated 
electronic systems, such as copper oxides or heavy fermion systems, 
has attracted great interest in recent years. 
These observations have led to the investigation 
of the interplay between magnetic long-range order and 
novel quantum disordered states.
Particular attention has been paid to low-dimensional quantum 
spin systems in which the classical ground states are 
destabilized by strong quantum fluctuations,\cite{Fazekas,Gelfand1,Bhatt,Katoh,Troyer2}
which are experimentally realizable in several compounds.

The bilayer Heisenberg model 
$$
 H=J\left(\sum_{\langle i,j \rangle}
{\bf S}_{1,i}{\cdot}{\bf S}_{1,j}
+\sum_{\langle i,j \rangle}{\bf S}_{2,i}
{\cdot}{\bf S}_{2,j}\right)
+\sum_{i}{\bf S}_{1,i}{\cdot}{\bf S}_{2,i}\ , 
\eqno{(1.1)}
$$
(taking the interlayer exchange as the unit of energy,
and with $\langle i,j \rangle$ running over nearest neighbor
pairs on the square lattice)
has attracted considerable interest
\cite{Sandvik,Hida,Gelfand2,Weihong,Matsuda,Ng,Morr}
as a testing ground for
notions of quantum criticality.\cite{Chakravarty,Chubukov}
On increasing $J$ from zero, at $T=0$ the model passes from a
spin-disordered ground state through a critical point
to a N\'eel-ordered state --- the same phase diagram as
the nonlinear $\sigma$ model, with $1/J$ playing the role
of the coupling parameter $g$.

For $S=1/2$, this model is directly relevant to the
magnetic properties of several cuprate high-$T_c$ parent,
antiferromagnetic compounds (and possibly the high-$T_c$
materials themselves\cite{Millis}), and also to the spin-gap system
$\rm{BaCuSi_{2}O_{6}}$\cite{Sasago}. 

If an external field
$$
H_{\rm Zeeman}=-h\sum_i (S^z_{1,i}+S^z_{2,i})
$$
is applied to the bilayer Heisenberg model, new
$T=0$ phases and phase transitions are introduced.
In particular, if at $h=0$ the system is in the disordered,
gapped phase then there is a critical field $h_c$,
trivially related to the zero-field triplet gap $\Delta$,
since there is a simple Zeeman splitting of the
triplet elementary excitations.  
For $h>h_c$ the gap is closed and the system exhibits
algebraic order in the spin components transverse to the
applied field.

A great deal is known about this $T=0$ field-induced transition.\cite{Sachdev}
In particular, $d=2$ is the upper critical dimension, so critical
exponents take on their classical values (modulo log corrections)
in the bilayer Heisenberg model.  In all dimensions above the lower
critical dimension $d=1$, the dynamic exponent $z=2$ and the correlation
length exponent $\nu=1/2$.
A variety of detailed model studies for one-dimensional
spin-gap systems have been carried out.\cite{Affleck,Sakai,Hida2}
Work on two-dimensional systems exhibiting this transition
is sparser,\cite{ES2} and
what is particularly not well known 
is how high in temperature the associated quantum critical
region extends.
This is a matter of some significance, if one is to ascribe
the temperature dependence of properties of an experimental
system to quantum critical behavior.

In this paper, we study the field-induced transition
in the bilayer Heisenberg model by means
of bond-operator mean-field theory\cite{Bhatt}
and strong-coupling expansions\cite{Gelfand3,ES1}.
Our principal conclusion is that the quantum critical
regime is rather narrow, and that ``quantum-critical-like''
temperature dependences of quantities such as the
magnetization are not sufficient to demonstrate that
the system is in the true quantum critical regime.

Bond operator mean-field theory has been applied 
to the spin-ladder (1D) and weakly coupled spin-ladder (3D) models
\cite{Normand} and shown to yield quantitatively correct results.
Mean-field theory is a particularly appropriate tool for the
problem at hand because the model is at its upper critical dimension,
and so the asymptotic critical behavior will be correctly
described up to logarithms.
The strong-coupling expansions provide a good check of
the accuracy of the mean-field theory, in the 
regimes where the former are reliable.

In \S{2}, we describe the bond-operator mean-field theory,
where the bilayer Heisenberg model is described in terms of 
three types of interacting bosons with a local constraint. 
In \S{3}, ground state phase diagram in the $h$-$J$ plane 
and the critical properties of phase transitions 
are discussed. In \S{4}, thermodynamic properties of 
quantum disordered and quantum critical phases in a magnetic field 
are presented. 
The last section (\S{5}) offers a summary and discussions. 

\section{Bond-Operator Mean-Field Theory}

Let us consider the quantum disordered phase of (1.1) 
in the limit $J\ll 1$ which is 
the analytic continuation of the interlayer dimer Hamiltonian ($J=0$). 
Then we describe the phase by 
introducing bosonic degrees of freedom corresponding to 
local dimer states (singlet and triplet states) following the 
bond-operator representation of Sachdev and Bhatt\cite{Bhatt}: 
$$
|s>=s^{\dagger}|0>=\frac{1}{\sqrt{2}}(|\uparrow\downarrow >
-|\downarrow\uparrow >)
\eqno{(2.1)}
$$
$$
|t_x>=t_x^{\dagger}|0>=-\frac{1}{\sqrt{2}}(|\uparrow\uparrow >
-|\downarrow\downarrow >)
\eqno{(2.2)}
$$
$$
|t_y>=t_y^{\dagger}|0>=\frac{i}{\sqrt{2}}(|\uparrow\uparrow >
+|\downarrow\downarrow >)
\eqno{(2.3)}
$$
$$
|t_z>=t_z^{\dagger}|0>=\frac{1}{\sqrt{2}}(|\uparrow\downarrow >
+|\downarrow\uparrow >) . 
\eqno{(2.4)}
$$
Here the four types of bosons satisfy bosonic commutation 
relations, and the left and right arrows in the kets represent 
the spin states on the first and second layers respectively. 
However there is a significant difference from usual bosonic systems 
because in the physical subspace only a single boson is allowed on each 
dimers. Thus the  Hilbert space is restricted by a 
local constraint on the number operators of these bosons 
in each dimer 
$$
   s^{\dagger}_{i}s_{i}
+t^{\dagger}_{\alpha ,i}t_{\alpha ,i}=1 \, 
\eqno{(2.5)}
$$
where the summation convention is used for Greek indices.

In terms of these bosons, the spin operators in each layer can be expressed as 
$$
  S_{1,i}^{\alpha}=\frac{1}{2}(s^{\dagger}_{i}t_{\alpha ,i}+
t^{\dagger}_{\alpha ,i}s_{i}-i{\epsilon}_{\alpha\beta\gamma}
t^{\dagger}_{\beta ,i}t_{\gamma ,i})  
\eqno{(2.6)}
$$
$$
  S_{2,i}^{\alpha}=\frac{1}{2}(-s^{\dagger}_{i}t_{\alpha ,i}-
t^{\dagger}_{\alpha ,i}s_{i}-i{\epsilon}_{\alpha\beta\gamma}
t^{\dagger}_{\beta ,i}t_{\gamma ,i}) .  
\eqno{(2.7)} 
$$
Then the bond-operator expression of Hamiltonian (1.1) is obtained 
in the form 
$$
  H=\sum_i (-\frac{3}{4}s_{i}^{\dagger}s_{i}+\frac{1}{4}
t^{\dagger}_{\alpha ,i}t_{\alpha ,i})
-\sum_i {\mu}_{i}(s^{\dagger}_{i}s_{i}+
t^{\dagger}_{\alpha ,i}t_{\alpha ,i}-1) 
$$
$$
+\frac{J}{2}\sum_{\langle i,j\rangle}(s^{\dagger}_{i}{s^{\dagger}}_{j}t_{\alpha ,i}t_{\alpha ,j}+  
s^{\dagger}_{i}s_{j}t_{\alpha ,i}t^{\dagger}_{\alpha ,j}
+\rm h.c.)
$$
$$
-\frac{J}{2}\sum_{\langle i,j\rangle}(1-{\delta}_{\alpha\beta})
(t^{\dagger}_{\alpha ,i}t_{\beta ,j}t^{\dagger}_{\alpha ,i}
t_{\beta ,j}-
t^{\dagger}_{\alpha ,i}t_{\beta ,j}t^{\dagger}_{\beta ,i}
t_{\alpha ,j}) \ .
\eqno{(2.8)}
$$
Here we have introduced local chemical potentials $\mu_i$ 
to account for the local constraint (2.5). 
Mean-field theory of the disordered phase is constructed 
by the following procedures.
(1) An average chemical potential $\mu\equiv\mu_i$ is 
introduced to approximate the local constraints 
by a global constraint.
(2) Bose-Einstein condensation of singlet states is assumed:
$\langle s^{\dagger}_{i}\rangle=\langle s_{i}\rangle=s$, 
where $\langle \cdots\rangle $ denotes 
the thermodynamic average.
(3) Terms containing four triplet operators are dropped. 
Here we would like to comment on the last numerically simplified 
assumption. The quadratic terms in $t_\alpha$-operators will 
in tern lead to a nonzero expectation value of 
$\langle t_{\alpha}t_{\alpha}\rangle$. 
By taking quadratic decouplings of fourth-order terms in $t_\alpha$, 
we found that the inclusion of these terms changes 
the numerical results only slightly, 
even near the critical coupling 
(for example, this lowers $J_c$ discussed in the next 
section about $5\%$). 
 
The resulting mean-field Hamiltonian is easily diagonalized 
using the Bogoliubov transformation: 
$$  
 \xi_{k\alpha}=\cosh\theta_{k}t_{k\alpha}
+\sinh\theta_{k}t^{\dagger}_{-k\alpha}
\eqno{(2.9)}
$$
$$
 \cosh^{2}\theta_{k}=\frac{1}{2}(\Lambda_{k}/\omega_{k}+1) 
\eqno{(2.10)}
$$
$$
 H_{MFT}(\mu ,s)=N(-\frac{3}{4}s^{2}-{\mu}s^{2}+\frac{5}{2}\mu
-\frac{3}{8}) +\sum_{k}\omega_{k}(\xi^{\dagger}_{k\alpha}
\xi_{k\alpha}+\frac{3}{2})
\eqno{(2.11)}
$$
$$
 \omega_{k}=\sqrt{ {\Lambda}^{2}_k-4{\Delta_k}^{2}}
\eqno{(2.12)}
$$
$$
 \Lambda_k = \frac{1}{4}-\mu+2Js^{2}\gamma_k \ , 
 \Delta_k = Js^{2}\gamma_k 
\eqno{(2.13)}
$$
$$ 
 \gamma_{k}=\frac{1}{2}(\cos{k_x}+\cos{k_y}) \ .
\eqno{(2.14)}
$$

The only effect of including an external field is to Zeeman-split
the excitations, so that instead of all three lying at $\omega_k$
they have energies $\omega_k$ and $\omega_k\pm h$.

The above mean-field description of the disordered phase 
will be valid if the temperature is much less than the 
minimum energy gap. However, at intermediate and 
higher temperatures the density 
of triplet bosons becomes sufficiently large that their interactions 
cannot be neglected. 
Troyer, Tsunetsugu and W\"{u}rtz\cite{Troyer} have
presented a simple prescription to modify the mean-field 
theory so that it correctly treats both the
low and high-temperature limits. 
They noticed that the main problem of the bosonic 
description is its overcounting of entropy 
due to the global constraint on the number of 
elementary excitations. 
They reweighted the $M$-boson part in the partition function 
so that each multiplet contributes the correct entropy. 
In their formulation, the free energy of the triplet bosons 
in a uniform magnetic field is expressed (per dimer) as
$$
  f=-\frac{1}{ \beta }\ln{ \{1+[1+2\cosh(\beta h)]z(\beta )\}}
\eqno{(2.15)}
$$
where
$$
 z(\beta )=\frac{1}{N}\sum_k \rm{e}^{-\beta\omega_k} \ .
\eqno{(2.16)}
$$
Using these formulae, one obtains self-consistent 
equations for the mean-field parameter $s$ and chemical 
potential $\mu$ by minimizing the total 
free energy ($F(s,\mu )=f(s,\mu)+\rm{constant}(s,\mu)$) 
with respect to those parameters: 
$$ 
  -\frac{3}{2}-2\mu +\frac{1}{N}\sum_k 
J\frac{2\gamma_k}{\sqrt{1+2r\gamma_k } }
\left[3+\frac{2(1+2\cosh(\beta h))\rm{e}^{-\beta\omega_k}}
{1+[1+2\cosh(\beta h)]z(\beta )} \right]=0
\eqno{(2.17)}
$$
$$ 
  -s^{2}+\frac{5}{2}-\frac{1}{2N}\sum_k 
\frac{1+\gamma_k}{\sqrt{1+2r\gamma_k } }
\left[3+\frac{2(1+2\cosh(\beta h))\rm{e}^{-\beta\omega_k}}
{1+[1+2\cosh(\beta h)]z(\beta )} \right]=0
\eqno{(2.18)}
$$
where 
$$
r=2Js^{2}/(\frac{1}{4}-\mu ) \ .
\eqno{(2.19)}
$$
It is straightforward to verify that this approach leads to
the correct free energy and magnetization for isolated dimers ($J=0$)
for all temperatures and external fields.
The numerical analysis of the above equations when $J\not=0$
is discussed below. 

\section{Ground State Phase Diagram}

In the classical (large-$S$) limit, 
the Heisenberg bilayer with $h>0$ has three types of ground 
states. 
One is the fully polarized ferromagnet, realized 
in the high-field region. 
The others are canted phases 
in which the $z$ component of the magnetization is parallel 
to the external field to gain Zeeman energy, 
while $x$ and $y$ components of the spins are ordered to 
gain exchange energy. 
For quantum spins a disordered (dimer) phase appears around $J=0$ for
sufficiently small $h$. 
As mentioned in Introduction, 
the phase diagram of (1.1) in zero magnetic field
has been extensively studied. 
The order-disorder transition lies in the universality class 
of classical $d=3$ Heisenberg model, and is associated with
a dynamical exponent $z=1$. 
However, the
field-induced transitions\cite{Sachdev} have dynamical exponent $z=2$ and so
should lie in the universality class of a $d=4$ classical model. 
The reduction of symmetry from $O(3)$ to 
$U(1)$ due to the magnetic field leads one
to anticipate that the universality of the transition 
is $d=4$ $XY$ type. 

To estimate the phase boundary between dimer and canted phases, 
we performed $T=0$ strong-coupling expansions of the longitudinal 
susceptibilities transverse to the applied field 
up to the 8th order at the ordering vectors from 
interlayer dimer Hamiltonian (i.e. series expansions 
in powers of the intralayer couplings $J$ 
about the interlayer dimer singlet state) 
using connected cluster method\cite{Gelfand3}. 
By applying the differential approximant method\cite{Fisher} 
assuming power-law divergence of the susceptibilities 
at critical points for fixed values of the applied field $h$, 
$\chi\sim (J_{c}(h)-J )^{-\gamma}$, 
we obtained estimates for the critical lines $J_{c}(h)$ 
(and found associated critical exponents $\gamma$ close to the 
expected value of 1). 
We could also estimate phase boundaries using the triplet excitation gaps 
at zero field which had been studied 
by strong-coupling expansions\cite{Gelfand2} previously. 
These different estimates of the critical lines are consistent 
with each other, as shown in Fig.~1. 
The boundaries between the fully polarized phase 
and canted phases shown in that figure
can be obtained from the lowest 
one magnon excitation gap in linear spin-wave theory.

Next we compare the results of the bond-operator 
mean-field theory at $T=0$ and $h=0$ with 
the results from the series expansions. 
At $T=0$, the self-consistent equations 
(2.17), (2.18) for the two parameters $s$ and $\mu$ can be reduced to 
a single equation for the parameter $r$ introduced in (2.19),
namely, 
$$
  r=J(5-3K(r))
\eqno{(3.1)}
$$
with
$$
 K(r)=\frac{1}{N}\sum_k \frac{1}{\sqrt{1+2r\gamma_k }} \ .
\eqno{(3.2)}
$$
Once the parameter $r$ is determined from the above set of 
equations, the mean-field parameter $s$ and chemical 
potential $\mu$ are obtained from
$$
  s^{2}=\frac{5}{2}-\frac{3}{4}(E(r)+K(r))
\eqno{(3.3)}
$$
and
$$
  \mu=-\frac{3}{4}+\frac{3}{2r}(E(r)-K(r))
\eqno{(3.4)}
$$
 where
$$
 E(r)=\frac{1}{N}\sum_k \sqrt{1+2r\gamma_k } \ .
\eqno{(3.5)}
$$ 

As $J$ grows from 0, the triplet gap at the wave vector $k=(\pi,\pi)$ 
decreases, and the gap vanishes at $J=J_c$, signaling 
the instability of dimer phase. 
The critical value $J_c$ can be obtained by setting $r=1/2$ 
and $J=J_c$ in (3.1), giving the value 
$J_c=0.437$ which is somewhat larger than 
the series expansion estimate 0.393. 
In addition we obtain a critical spin-wave velocity 
0.715 which is somewhat smaller than 
the series expansion estimate 0.744. 
The triplet gap for various values of $J$ in the dimer phase,
and excitation spectra for $J=0.2$ and 0.35, were obtained from 
numerical solutions of the mean-field equations, with
the results shown in Figs.~2 and 3. 
For negative $J$, the mean-field theory gives 
the same critical parameters, changing only the sign of $J_c$. 
For comparison, the series expansion estimate for the
negative critical coupling is\cite{Matsushita} $-0.433$,
so mean field theory is remarkably accurate in that case.

In the following section, we restrict our discussion of 
several thermodynamic properties only for positive $J$ 
but the qualitative results of thermodynamic behavior 
are the same for negative $J$. 

\section{Finite Temperature Properties}

Our principal purpose in this section is to discuss 
the finite temperature quantum critical properties 
of the bilayer system, where a magnetic field 
is the tuning parameter and the system has a quantum 
disordered ground state at zero field. 
We analyzed these properties on the basis of the bond-operator 
mean-field theory as well as finite-temperature
strong-coupling expansions up to 
5th order (i.e. series expansions of the 
free energy in powers of the intralayer coupling $J$) using 
connected cluster method\cite{Haaf}. 
Calculations of the latter variety for some parameter cases 
have already been carried out to 
8th order by Elstner and Singh for bilayers both with\cite{ES2} and 
without\cite{ES1} an external field. 
It turns out that the results from directly summing 5th order 
series are not very different from those of 8th order series,
at least in the regimes where we expect these calculations to be
most reliable. 

Let us first discuss the thermodynamic properties 
of the system when $h=0$. 
In Fig.~4, we show the energy gap $\Delta$ of triplet 
excitations as a function of temperature as
obtained from the mean-field theory. 
For values of $J$ less than $J_c$, $\Delta$ is finite at $T=0$ and 
shows a sharp rise with increasing temperature
and eventually tends to saturate, reflecting a rapid extinction 
of long-range correlations. However, at $J=J_c$, $\Delta$ 
rises from zero linearly with temperature, 
consistent with the prediction of the non-linear $\sigma$ model,
$\Delta\simeq 1.04T$\cite{Chubukov}. 
We have calculated the specific heat $C_V$ 
and uniform susceptibility $\chi$ for $J=0.2$ 
as functions of temperature, using both mean-field
theory and series expansions, with the results shown 
in Fig.~5. 
Since the system has a quantum disordered ground state, 
the thermodynamic quantities show thermally activated 
behavior. 
For the specific heat, both methods yield nearly
identical results. For the susceptibility,
mean-field theory deviates by about $10\%$ 
at intermediate temperatures from the results
of series expansions. 

Let us now discuss the temperature dependence of
properties of the system at a field-induced transition. 
In Figs.~6 and 7, we show the 
magnetization and specific heat as functions of $T$ at the critical 
field $h=h_{c}=\Delta$ for $J=0.2$. 
In general, we expect power-law dependences
of thermodynamic quantities at low temperatures 
if the parameters of a system are set to a quantum critical point. 
Mean-field theory gives a linear $T$ dependence for both quantities 
in a very narrow low temperature region, $ T\leq 0.1 \ll \Delta$.
The agreement between two approaches is fairly good 
except for some artificial features seen in the summed series expansions
at low temperatures. 

The series expansions at finite temperature
should reduce to the strong-coupling expansion from 
the dimer singlet state in the limit $T=0$. 
Thus the thermodynamic quantities obtained by 
direct summation of finite 
series necessarily decrease exponentially at sufficiently low temperature. 
However we might have the possibility of  seeing 
quantum critical behavior at {\it intermediate\/} temperatures 
if the tuning parameter is set equal or 
close to the quantum critical point of the infinite system. 
Thereby we must be careful to judge whether or not 
the system is really exhibiting quantum critical behavior 
in the relevant temperature regime. 
In Fig.~6, it is found that the magnetization starts to show 
apparently linear behavior towards $T=0$ at $T\sim 0.3$,
but mean-field theory indicates that 
the system has not yet entered the quantum 
critical region, since at $T=0$ the magnetization
has a different slope than in the intermediate-$T$ ``linear'' regime. 
The specific heat behaves similarly, as seen in Fig.~7.
(The series expansion result
exhibits a pair of peaks and a dip 
but this should be regarded as an 
artifact due to low order of the expansion
and the crudeness of the extrapolation technique.)

The case we have presented above, for $J=0.2$, is actually the
one in which there is the best chance of observing the true
quantum critical regime for the field induced transition.
As $J\to0$ the quantum critical regime vanishes since there 
is no way that the critical properties will be exhibited
for $T>J$.  As $J\to J_c$, $h_c\to0$ and it is natural to
expect that for $T>h_c$ (but not too large)
the system will exhibit the quantum critical behavior
associate with the {\it zero-field\/} transition.
Thus the field-induced transition's quantum critical regime is squeezed
at both ends.  We have made rough estimates of the extent
of that regime as a function of $J$, based on the behavior of
$M(T)$, with the results shown in Fig.~8.  It therefore appears
that observing the quantum critical regime for the field-induced
transition will, in general, be very difficult to achieve either
in numerical simulations or experiments.

\section{Summary and Discussion}

We have investigated the ground state phase 
diagram and thermodynamic properties 
of bilayer Heisenberg model on the square-lattice 
in a uniform magnetic field. 
The ground state phase diagram of the model has been 
investigated by means of strong-coupling expansion around
the dimer limit (considering both the gap and
longitudinal susceptibility) and linear spin-wave theory. 
We found that the longitudinal susceptibility exponent 
for the field-induced transition appears to be very close to 
$z\nu=1$, as expected.  
However, we cannot exclude the possibility of logarithmic 
corrections to scaling expected at upper critical 
dimensions, and further field-theoretical investigations 
\cite{Sachdev2} are needed to clarify this point. 
It might also be interesting to calculate the transverse
susceptibility, for which one expects an exponent of
$(z+d)\nu=2$.

Thermodynamic properties of the model
were investigated also by means of 
bond-operator mean-field theory, 
and its results were compared to those of the finite temperature 
strong-coupling expansions. 
We found the theory provides a simple but reasonably good 
description of thermodynamic quantities at all temperatures. 
It appears that the quantum critical region for
field-induced transitions in this system is restricted 
to extremely low temperatures, even though the magnetization 
exhibits what might appear to be the expected quantum
critical power law behavior at higher temperatures. 

Finally, we mention that a transition entirely analogous 
to the field induced transition in Heisenberg bilayers 
can take place in bilayer quantum Hall systems.\cite{Zheng} 
Recently, Troyer and Sachdev\cite{Troyer3} presented quantum Montecalro 
simulation on Heisenberg bilayers in a magnetic field 
to determine the {\it universal} Kosterlitz-Thouless transition 
temperature $T_{KT}$ in the vicinity of a zero-field 
quantum critical point. 
The universality of $T_{KT}$ when the ground state is ordered 
leads us to an anticipation of the universality of 
a quantum critical crossover temperature 
when the ground state is disordered ($h<h_c=\Delta$) 
(which is not our estimated one but the one 
between field-induced quantum critical and 
quantum disordered regions). 
We leave this issue to a future study. 

\vspace{1cm}

\section*{Acknowledgements}
This work has been supported by the U. S. National Science
Foundation through grant DMR 94--57928 (MPG).

\newpage
\section*{FIGURE CAPTIONS}
Fig.~1. \hspace{6pt} Phase diagram of bilayer 
Heisenberg model in the magnetic field. 
The canted phases are specified by the ordering vectors 
$(\pi,\pi;\pi)$ and $(0,0;\pi)$, where the third component 
$\pi$ indicates antiferromagnetic orientation between layers. 
The open circles are series expansion estimates 
of the triplet excitation gaps\cite{Gelfand2}. 
\newline
Fig.~2. \hspace{6pt} Triplet gap as a function of $J(>0)$. 
The dots and solid line are series expansion estimates
\cite{Gelfand2} and mean-field theory results, respectively. 
\newline
Fig.~3. \hspace{6pt} Triplet excitation spectra for $J=0.2, 0.35$. 
The dots and lines are series expansion estimates and mean-field theory results, respectively. 
\newline 
Fig.~4. \hspace{6pt} Triplet gaps for various $J\,(\leq J_c)$ 
as a function of temperature $T$. 
\newline
Fig.~5. \hspace{6pt} Uniform susceptibility and specific heat 
(per a dimer) as a function of temperature for $J=0.2$ and $h=0$. 
The dots and solid lines are the results of 
series expansion up to the 5-th order and mean-field theory, 
respectively. 
\newline
Fig.~6. \hspace{6pt} Magnetization at the critical field for $J=0.2$
as a function of temperature. 
The symbols are the same as in Fig.~5. 
The broken line is just a guide to the eye. 
\newline
Fig.~7. \hspace{6pt} Specific heat at the critical field  for $J=0.2$
as a function of temperature. 
The symbols are the same as in Fig.~5. 
The broken line is just a guide to the eye. 
\newline
Fig.~8.\hspace{6pt} Estimates of the extent of the quantum critical
regime for the field-induced transition, as a function of $J$.
The curved line is just a guide to the eye.
\newline
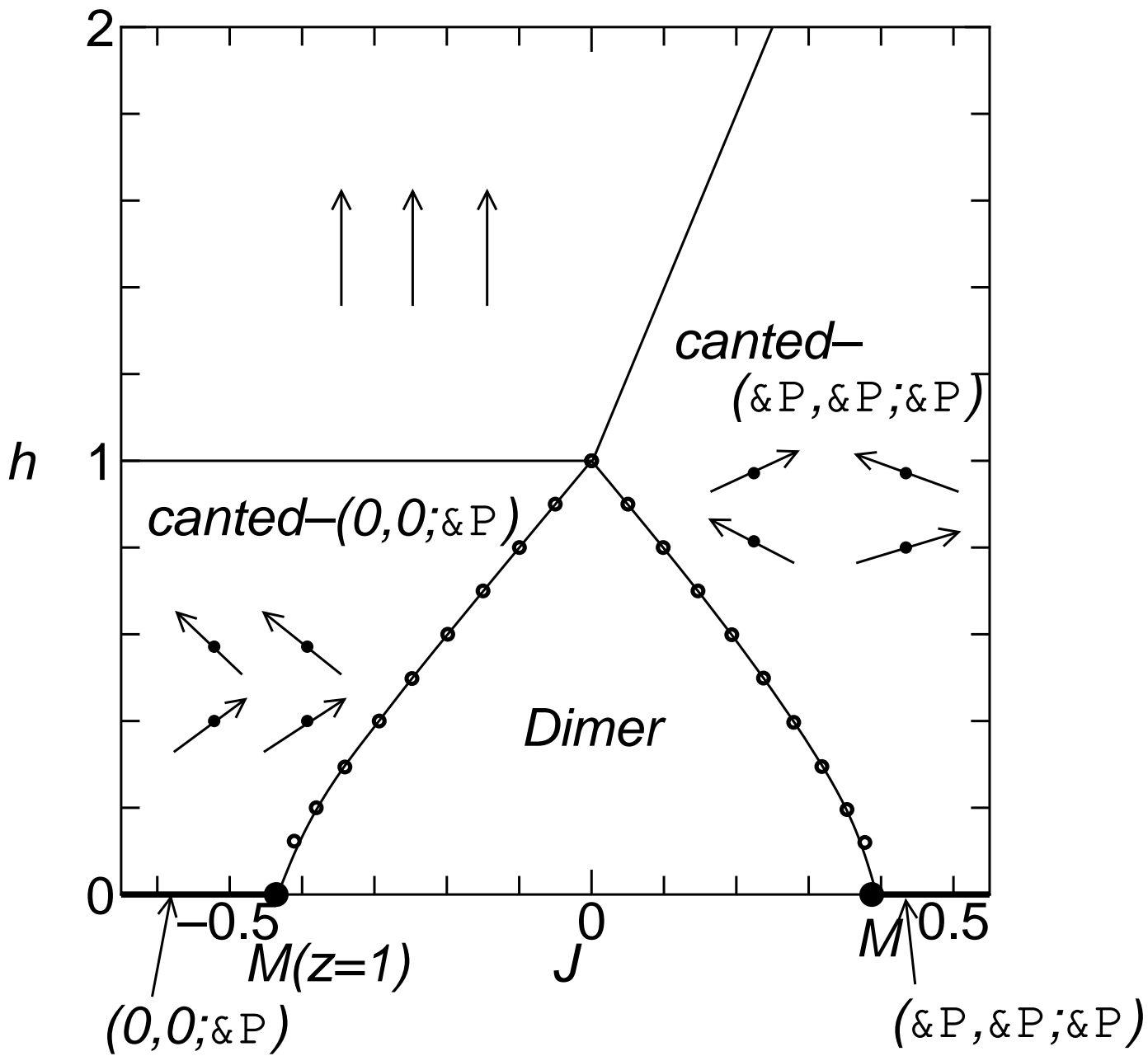
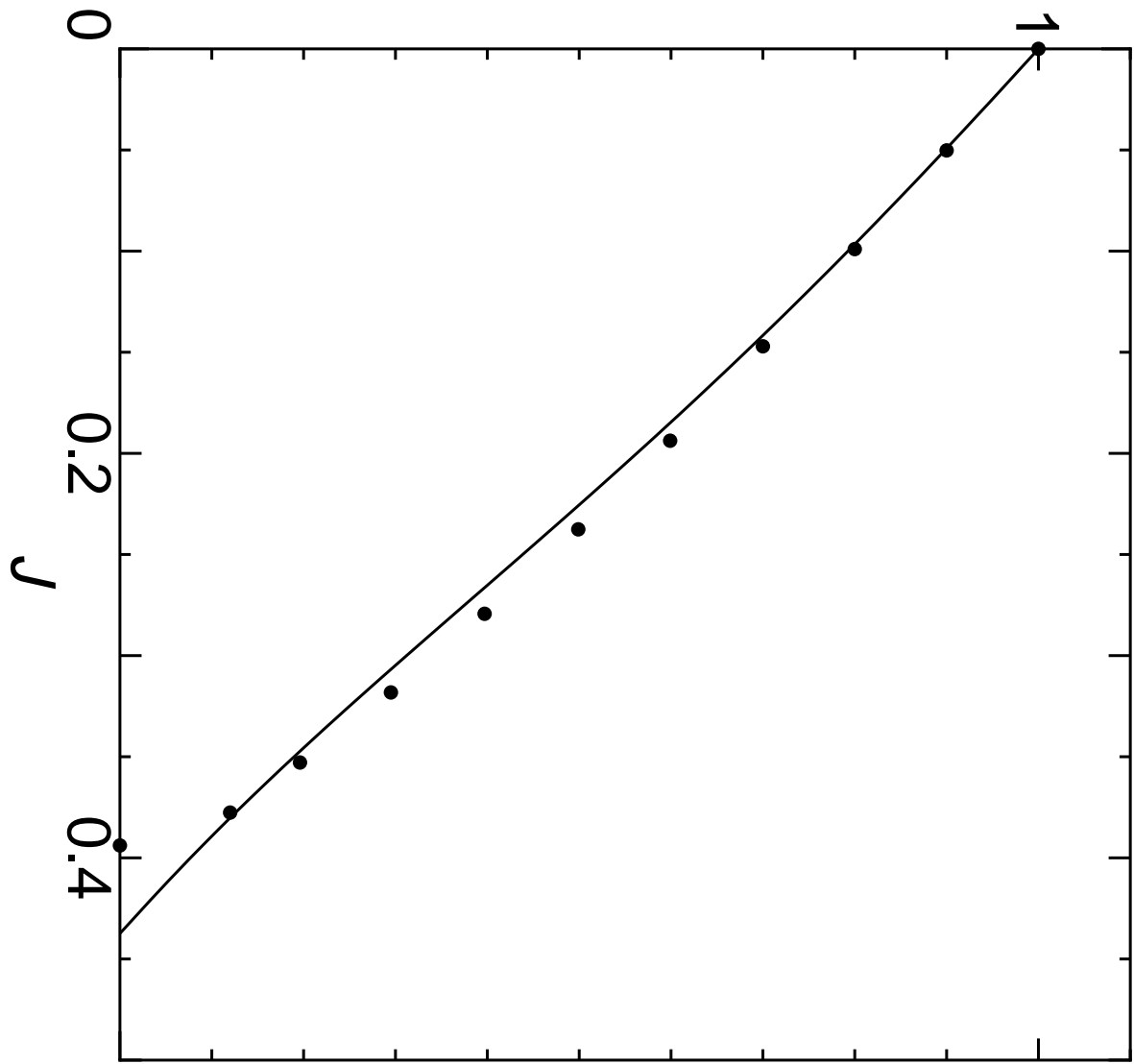
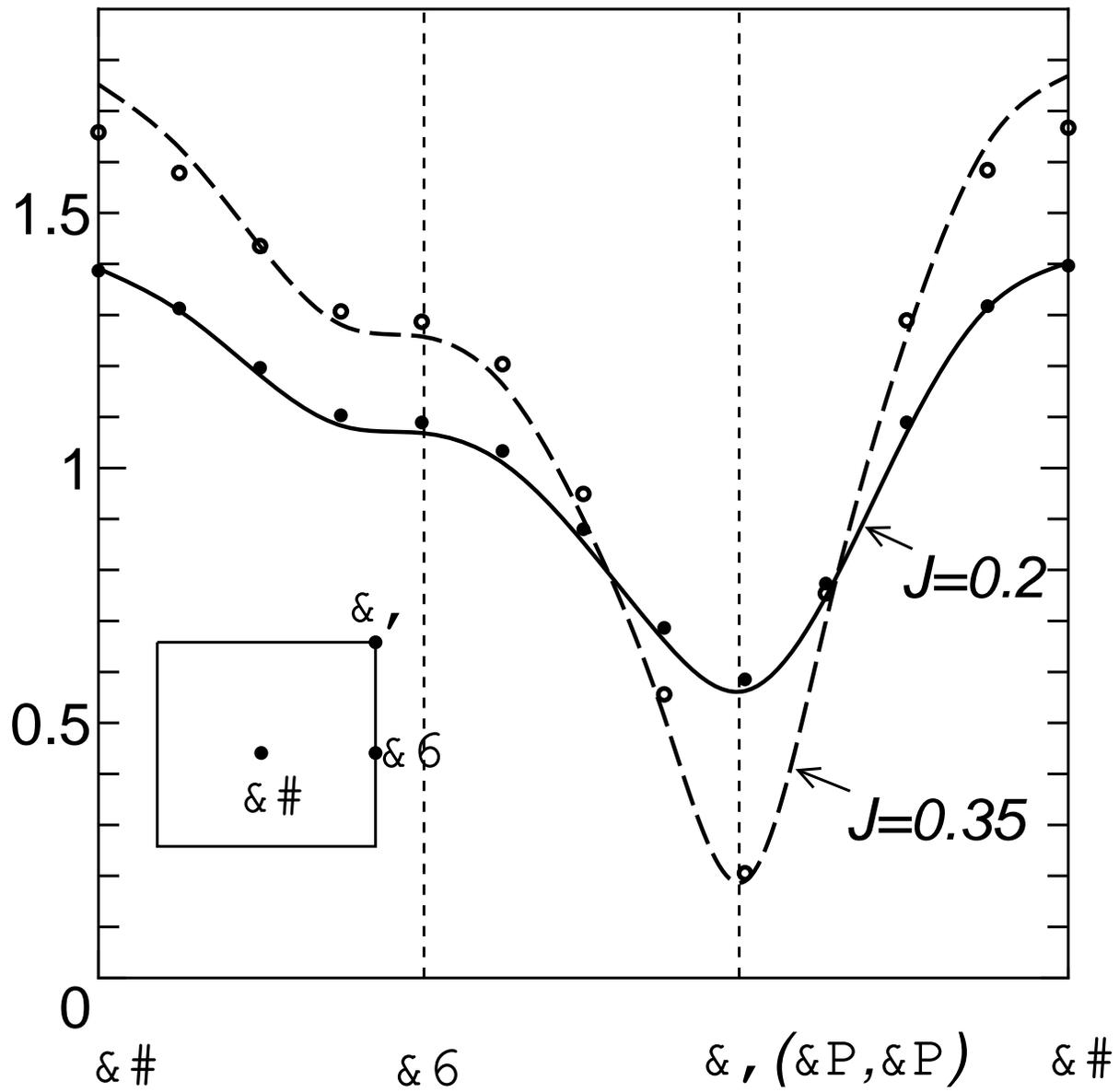
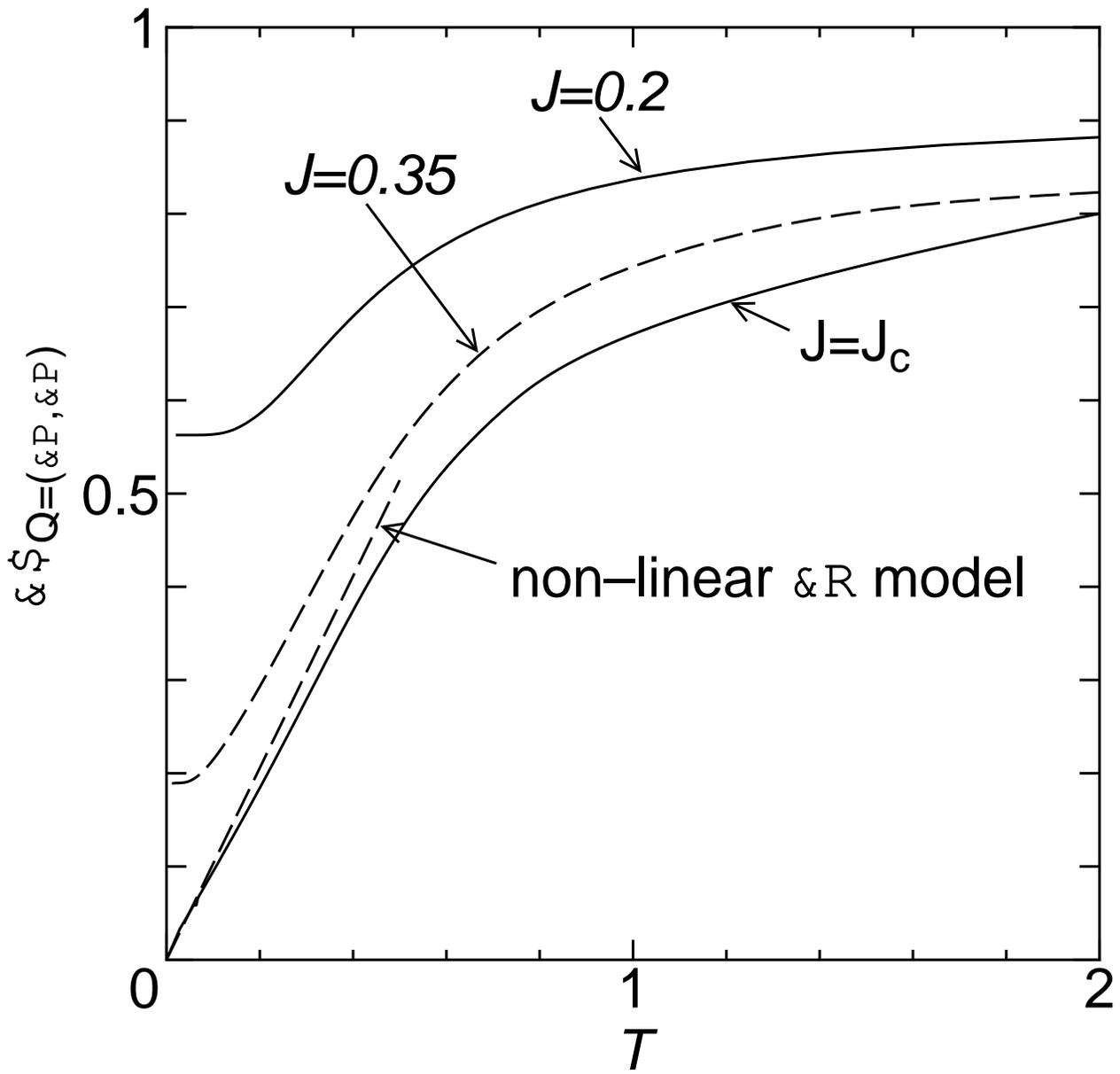
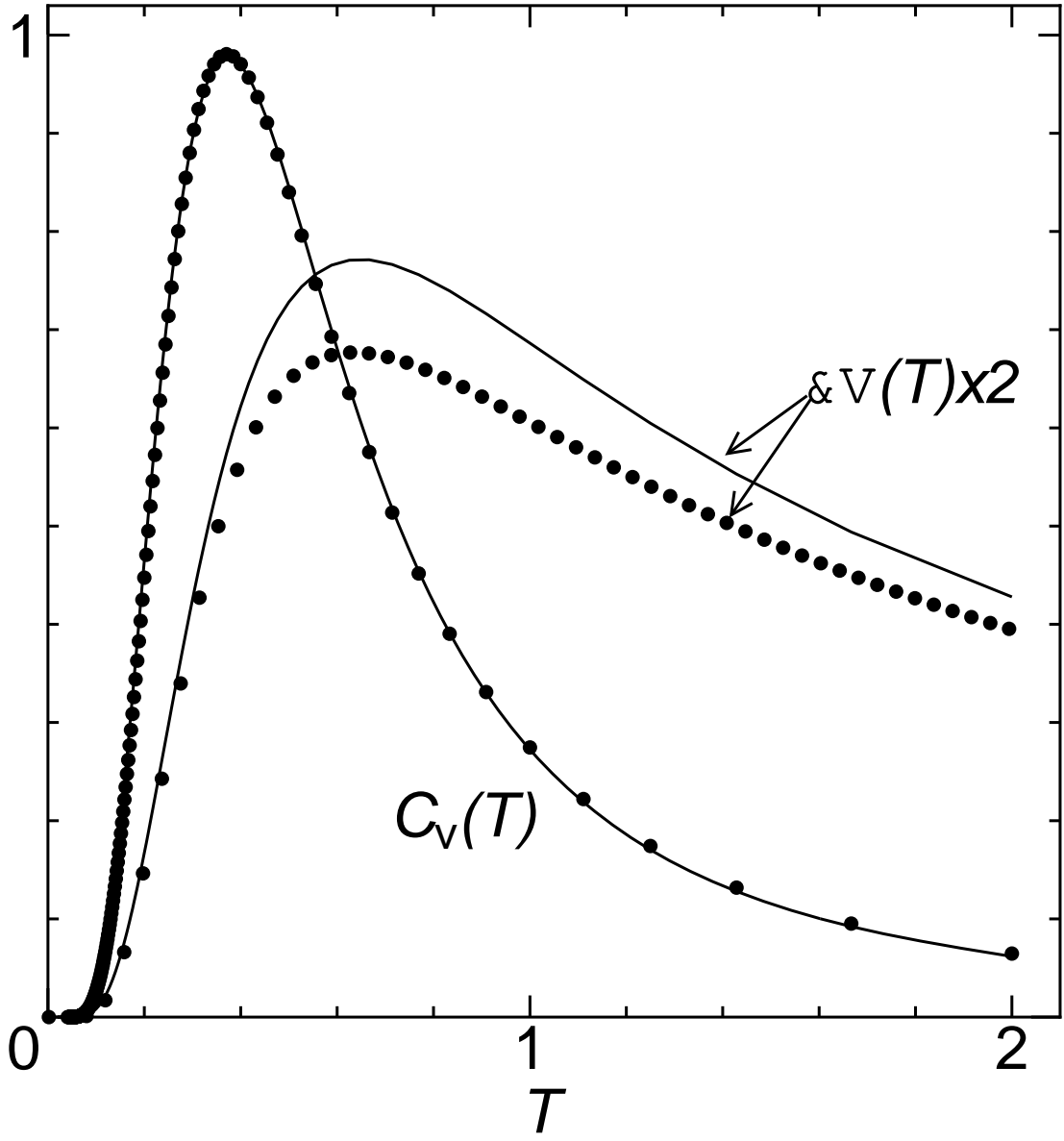
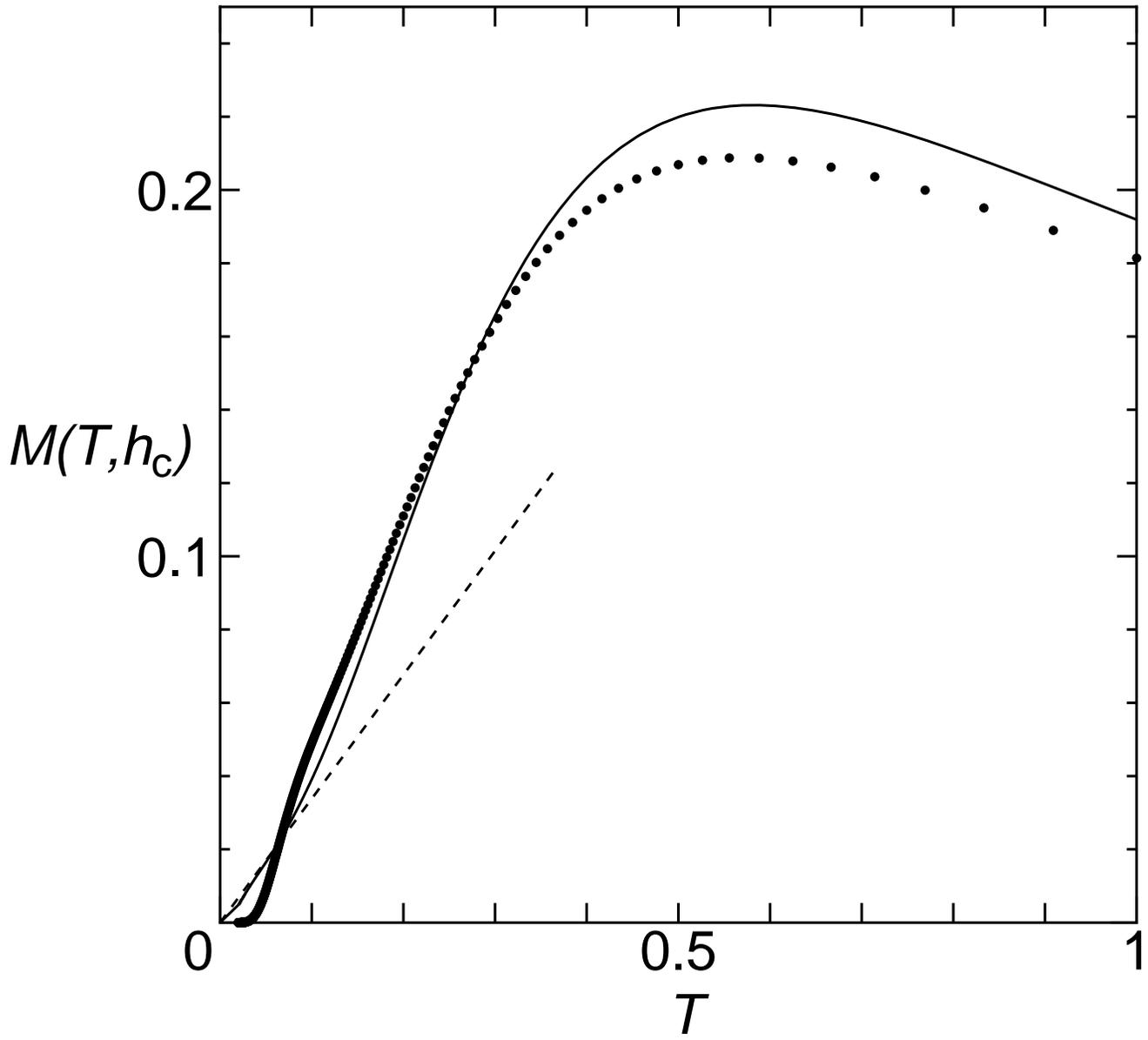
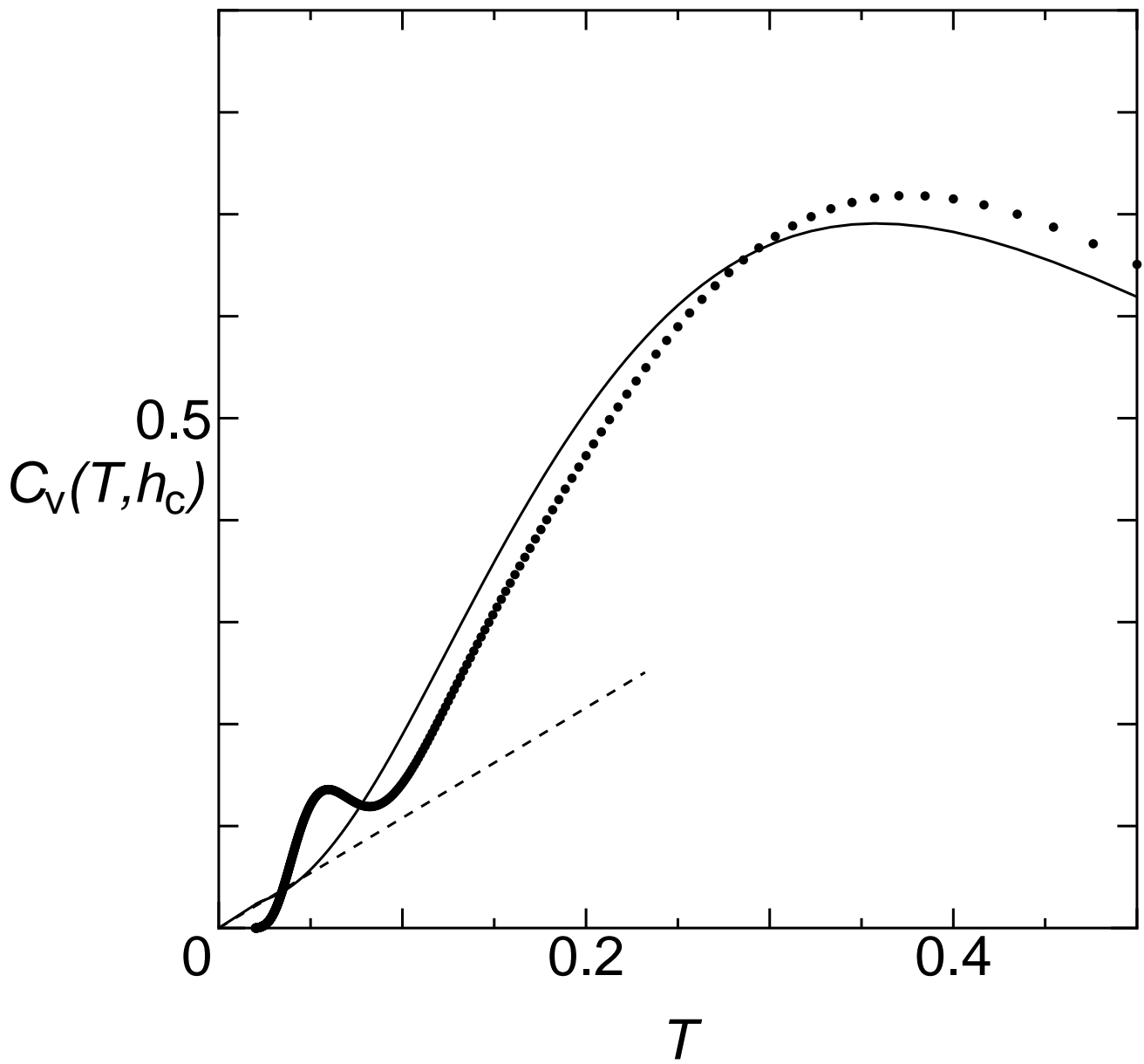
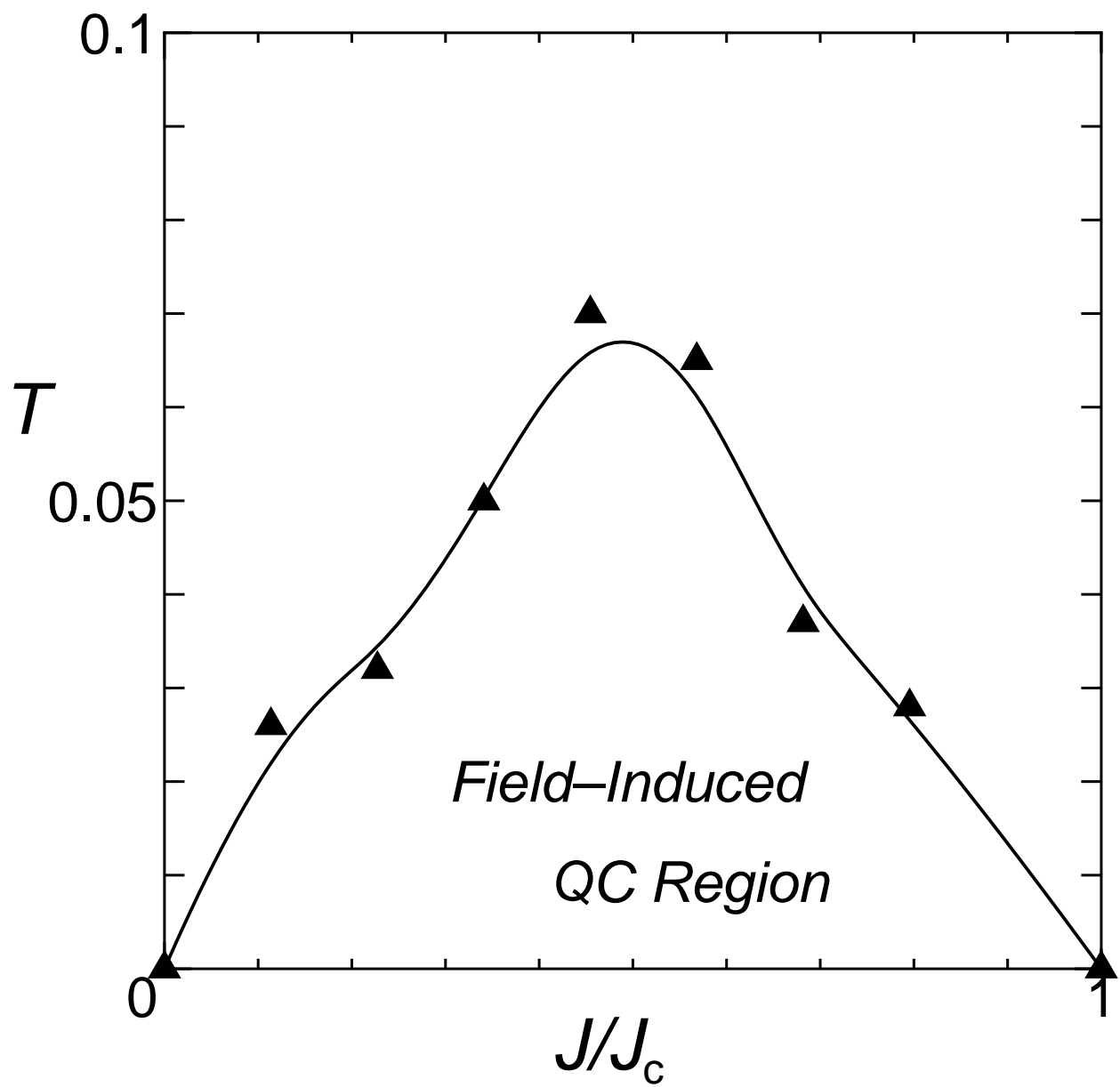
\end{document}